\documentclass[conference]{IEEEtran}

\usepackage{graphicx}
\usepackage{latexsym}
\usepackage{amsfonts}
\usepackage{amssymb}
\usepackage{psfrag}
\usepackage{cite}
\usepackage{subfigure}
\usepackage{comment,cancel,float}
\usepackage{amsmath,epsfig,epstopdf,color,lineno,makeidx,booktabs,multirow}
\usepackage[normalem]{ulem}
\usepackage{stfloats}

\begin{document}

\title{Efficient Symbol Detection for the FSO IM/DD System with Automatic and Adaptive Threshold Adjustment: The Multi-level PAM Case}

\author{ \IEEEauthorblockN{Tianyu Song and Pooi-Yuen Kam}
\IEEEauthorblockA{Department of Electrical and Computer Engineering\\
National University of Singapore, Singapore 117583\\
Email: \{song.tianyu, elekampy\}@nus.edu.sg}
}

\markboth{IEEE Photonics Journal}{Volume}

\maketitle

\begin{abstract}
To detect $M$-ary pulse amplitude modulation signals reliably in an FSO communication system, the receiver requires accurate knowledge about the instantaneous channel attenuation on the signal.  
We derive here an optimum, symbol-by-symbol receiver that jointly estimates the attenuation with the help of past detected data symbols and detects the data symbols accordingly. 
Few pilot symbols are required, resulting in high spectral efficiency. 
Detection can be performed with a very low complexity.
From both theoretical analysis and simulation, we show that as the number of the detected data symbols used for estimating the channel attenuation increases, the bit error probability of our receiver approaches that of detection with perfect channel knowledge.  
\end{abstract}


\section{Introduction}

\begin{figure}
\centering
\subfigure[Transmitter Constellation]{
\includegraphics{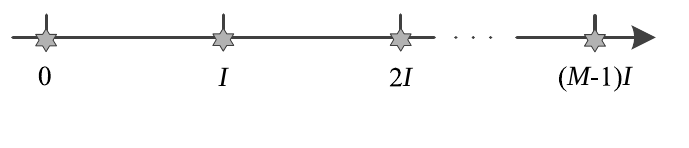}
\label{fg:cons_Tx}
 }
 \subfigure[Receiver Constellation]{
\includegraphics{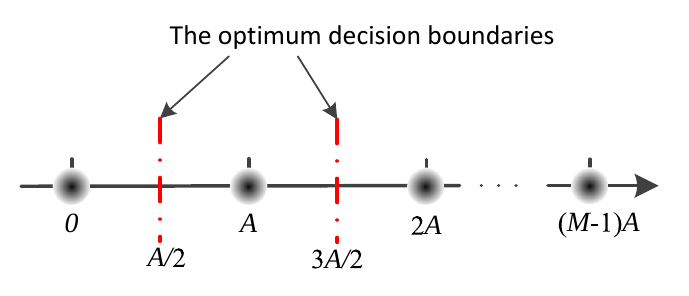}
\label{fg:cons_Rx}
}
\caption{Constellations}
\end{figure}

\begin{figure}
\centering
\includegraphics{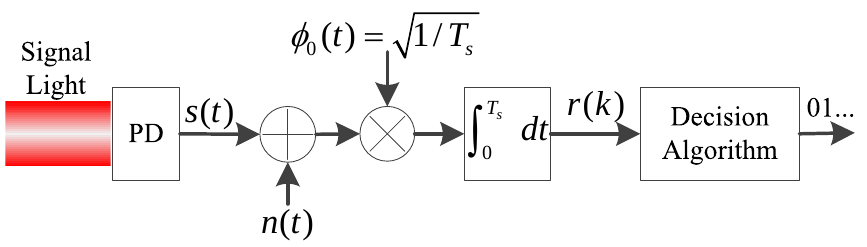}
\caption{Receiver diagram}
\label{fg:receiver}
\vspace{-20pt}
\end{figure}

In most current free space optical (FSO) communication systems, for reasons of simplicity, intensity modulation with direct detection (IM/DD) is used. 
To improve the spectral efficiency, we consider $M$-ary pulse amplitude modulation (MPAM) here by presuming that the receiver photodetector will not be saturated by the impinging optical signal level that has the highest energy. 
The transmitter side constellation is shown in Fig. \ref{fg:cons_Tx}, where $I$ denotes the minimum signal intensity distance and the transmitter average power is defined as $\bar{P} =\frac{1}{M} \sum_{j=0}^{M-1}jI$.  
Inside the receiver, as shown in Fig. \ref{fg:receiver}, there is an integrator that integrates the photo current for each symbol period $T_s$.
For the $k$th symbol interval ($(k-1)T_s,kT_s$), the received electrical signal $r(k)$ is obtained by sampling the integrator at time $t = kT_s$.
Since the photo current can be assumed to be constant during the integration time, $r(k)$ can be expressed as \cite{Zhu2002FSOComm}
\begin{align}
r(k) = \sqrt{{1}/{T_s}}RhIm(k)T_s+n(k),
\label{eq:signal1}
\end{align} where $R$ is the responsivity of the photo detector and $h$ denotes the instantaneous channel gain.
The transmitted data symbol $m(k)$ takes on any value from set $\{0, 1, ... ,M-1\}$ with equal probability, and Gray mapping of bits onto the levels is assumed. 
In practice, multiplying the received signal by the normalising basis $\phi_0(t)=\sqrt{{1}/{T_s}}$, which is shown in Fig. \ref{fg:receiver}, is not necessary. 
We use it here because we want to simplify later performance analysis by normalizing the discrete additive white Gaussian noise (AWGN) term $n(k)$ such that $\mathbb{E}[n(i)n(j)]=\delta _{ij} {N_0}/{2} $ conditioned on that the continuous noise term $n(t)$ is an additive white Gaussian random process with mean zero and two-sided spectral density $N_0/2$.
By defining $A=\sqrt{{1}/{T_s}}RhIT_s$ as the instantaneous receiver-side electrical-domain minimum signal distance, $r(k)$ can be modelled as  \cite{Song2014GlobeCom}
\begin{align}
r(k)=  Am(k)+n(k).
\end{align}
Correspondingly, the receiver side constellation is shown in Fig. \ref{fg:cons_Rx}. 
Since atmospheric turbulence and pointing errors cause fluctuations in the intensity of the received signal, i.e., $h$ is time-varying, therefore, $A$ is time-varying and is modelled as $A=2dh$ where $2d$ denotes the minimum signal distance when $h=1$, i.e., $2d=\sqrt{{1}/{T_s}}RIT_s$, which is related to the average transmit power $\bar P$, i.e., 
\begin{align}
2d =\sqrt{{1}/{T_s}}RIT_s = \frac{2\sqrt{T_s}R\bar{P}}{M-1}.
\label{eq:distance}
\end{align} 

For the on-off keying (OOK) system that is considered as the simplest MPAM ($M=2$) case, the turbulence and pointing error induced fading is the major issue that degrades system performance.
Thus, the accurate channel state information (CSI), i.e., the accurate instantaneous value of $h$, is required for properly adjusting the decision threshold. 
In previous studies,  frequent insertions of pilot symbols \cite{Zhu2002PSAM} or co-propagating reference light \cite{YuChangyuan2012performanceOE} is used to estimate the CSI. 
In this paper, when it comes to pilot symbols, we specifically refer to pilot symbols used for channel estimation.
In practice, pilot symbols are also required for other purposes such as timing synchronization.  
Since pilot symbols do not carry data,  the use of them results in a system spectral efficiency reduction and additional energy overheads.
Therefore, we hope to minimize the amount of pilot symbols. 
The co-propagating reference light has the same problem, and besides, since it has a different frequency with the signal-bearing light, the estimation result may not be accurate. 
For MPAM systems ($M>2$), to detect signals reliably, the requirement of accurate CSI is more stringent. 
The difficulty of accurate CSI acquisition is the main reason that limits the use of MPAM in practical systems.
However, though these two above-mentioned methods can be extended to MPAM systems, they are not so suitable in practice.
The receiver design problem remains challenged by the fact that the CSI is hard to acquire at the receiver side if very few pilot symbols and no co-propagating reference light are used.

Since the channel coherence length $L_c$, defined as the number of consecutive data symbol intervals over which the channel gain $h$ can be considered to be constant, is a very large number ($>10^{4}$) for multi-Gbps systems\cite{Xie2011EffICC}, the data and the CSI can be jointly detected and estimated. 
In \cite{Song2014Arobust}, for the 2-level PAM system, i.e., the OOK system, based on the generalized likelihood ratio test (GLRT) principle, we derived a maximum likelihood (ML) sequence detection receiver (the GLRT-MLSD receiver) that jointly detects the data sequence and estimates the unknown channel gain. 
In \cite{Song2014GlobeCom}, we have extended this GLRT-MLSD receiver to multi-level PAM systems. 
However, as we discussed in \cite{Song2014GlobeCom}, the search complexity of the GLRT-MLSD receiver increases with the modulation order $M$ almost quadratically. 
Therefore, in section II of this paper, we derive a simpler decision-feedback (DFB), symbol-by-symbol receiver, whose implementation complexity is independent of the modulation order $M$. 
It uses the most recently detected data symbols to help estimate the instantaneous channel gain. 
As the number of the detected data symbols used to estimate the channel increases, the bit error probability (BEP) of this receiver approaches the Genie Bound, which is defined as the BEP of detection with perfect CSI (PCSI).

\section{The Decision-Feedback Receiver}
 
\subsection{Receiver Design}
We use $\hat m(k)$ to denote the decision on symbol $m(k)$ and $\hat{\mathbf{m}}(k,L)= [ \hat{m}(k-L+1)$, $\hat{m}(k-L+1)$, ..., $\hat{m}(k)]$ to denote the decision on the subsequence $\mathbf{m}(k,L) = [ {m}(k-L+1)$, ${m}(k-L+1)$, ..., ${m}(k)]$. 
The GLRT-MLSD receiver 
\begin{equation}
\hat{\mathbf{m}}(k,L) =\arg \underset{\mathbf{m}(k,L)} \max \ \lambda(\mathbf m(k,L)),
\label{eq:GLRT_MLSD}
\end{equation}
which was derived in \cite{Song2014Arobust} for OOK systems and generalised in \cite{Song2014GlobeCom} for multi-level PAM systems, has been shown to be able to achieve the Genie Bound. 
In \eqref{eq:GLRT_MLSD}, $\lambda(\mathbf m(k,L))$ is the decision metric and is given as 
\begin{equation}
  \lambda(\mathbf m(k,L)) =    \frac{(\mathbf{r}(k,L)\cdot\mathbf{m}(k,L))^2}{\|\mathbf{m}(k,L)\|^2},
\label{eq:GLRT_metric}
\end{equation}
where $\mathbf{r}(k,L)$ is the received signal subsequence with length $L$, i.e., $\mathbf{r}(k,L) = [ {r}(k-L+1)$, ${r}(k-L+1)$, ..., ${r}(k)]$.
This GLRT-MLSD receiver continuously and implicitly performs ML estimation of the $h$ by \cite[Eq. (16)]{Song2014GlobeCom}.
Here, for any hypothesized subsequence $\mathbf m(k,L)$, $A=2dh$ can be estimated similarly by performing 
\begin{equation}
\hat A(\mathbf m(k,L))=\frac{\mathbf{r}(k,L)\cdot\mathbf{m}(k,L)}{\|\mathbf{m}(k,L)\|^2} .
\label{eq:A_estimate}
\end{equation}

Our GLRT-MLSD receiver has to evaluate the decision metrics of all possible subsequences
and choose the one with largest metric value as the decision.
In this section, based on \eqref{eq:GLRT_MLSD}, we develop a new simpler decision-feedback symbol-by-symbol receiver. 
We first assume that at each time $k$, all the past decisions before time $k$ have been completed, and  we use  $[\hat{\mathbf{m}}(k-1,L),m(k)]$ to denote the subsequence whose first $L$ elements are $\hat{m}(k-L)$, $\hat{m}(k-L+1)$, ..., $\hat{m}(k-1)$ and last element is $m(k)$.
Thus, at time $k$, the detection rule given in \eqref{eq:GLRT_MLSD} can be considered as choosing the $m(k)$ that maximizes the decision metric value of subsequence $[\hat{\mathbf{m}}(k-1,L),m(k)]$, i.e., 
\begin{equation}
\hat{m}(k) =  \arg \max_{m(k)} \lambda ([\hat{\mathbf{m}}(k-1,L),m(k)]).
\label{eq:dfb_0}
\end{equation} 
Since $[\mathbf{r}(k-1,L),r(k)]=\mathbf{r}(k,L+1)$ is independent of the detection result, after substituting \eqref{eq:GLRT_metric} into \eqref{eq:dfb_0}, \eqref{eq:dfb_0} is equivalent to
 \begin{align}
& \hat{m}(k) \nonumber \\ 
  = & \arg \min_{m(k)}   \big( \|[\mathbf{r}(k-1,L),r(k)]\|^2   
 -  \lambda ([\hat{\mathbf{m}}(k-1,L),m(k)]) \big)
   \nonumber \\
  = & \arg \min_{m(k)}  \| [\mathbf{r}(k-1,L), r(k)]   \nonumber \\
 & \qquad   - \hat{A}([\hat{\mathbf{m}}(k-1,L),m(k)]) [\hat{\mathbf{m}}(k-1,L),m(k)]  \|^2 .
\label{eq:dfb_deriviation0}
\end{align}

As discussed in section 2, the channel coherence length $L_c$ is on the order of $10^4$, which is a very large number. 
The channel gain $h$ can be safely considered unchanged from time point $k-1$ to $k$. 
Thus, we can use the ML estimation on $A=2hd$ based on $\hat{\mathbf{m}}(k-1,L)$, denoted by  
\begin{align}
\hat{A}(\hat{\mathbf{m}}(k-1,L)) =  
\frac{\mathbf{r}(k-1,L) \cdot \hat{\mathbf{m}}(k-1,L)}{\|\hat{\mathbf{m}}(k-1,L)\|^2}, 
\label{eq:ch_est_b4sss}
\end{align}
to approximate $\hat{A}([\hat{\mathbf{m}}(k-1,L),m(k)])$, with very high accuracy. 
This enables us to further simplify the decision rule \eqref{eq:dfb_deriviation0} to \eqref{eq:dfb_deriviation2}. 
\begin{figure*}
\begin{align}
& \hat{m}(k)  =\arg \min_{m(k)}   \left\| [\mathbf{r}(k-1,L), r(k)]  -  
\hat{A}(\hat{\mathbf{m}}(k-1,L)) [\hat{\mathbf{m}}(k-1,L),m(k)] \right\|^2   \nonumber \\ 
 = & \arg \min_{m(k)}  \left( \left\|  \mathbf{r}(k-1,L)  - \hat{A}(\hat{\mathbf{m}}(k-1,L)) \hat{\mathbf{m}}(k-1,L) \right\|^2  +   \left\|  r(k)  -  \hat{A}(\hat{\mathbf{m}}(k-1,L)) m(k) \right\|^2 \right).
 \label{eq:dfb_deriviation2}
\end{align}
\vspace{-30pt}
\end{figure*}
Since term $\left\|  \mathbf{r}(k-1,L)  - \hat{A}(\hat{\mathbf{m}}(k-1,L)) \hat{\mathbf{m}}(k-1,L) \right\|^2 $ is independent of $m(k)$, we eliminate it and simplify the decision rule as 
\begin{equation}
\hat{m}(k) =  \arg \min_{m(k)}   ( r(k)  -  \hat{A}(\hat{\mathbf{m}}(k-1,L)) m(k) ) ^2 .
\label{eq:dfb1}
\end{equation}
In principle, to decide based on \eqref{eq:dfb1}, one has to compute the value of $(r(k) - \hat A (\hat{\mathbf{m}}(k-1,L))m(k))^2$ for each   $m(k)\in\{0,1,2,...,M-1\}$ and then choose the $m(k)$ corresponding to the minimum $(r(k) - \hat A (\hat{\mathbf{m}}(k-1,L))m(k))^2$ value. 
Since $m(k)$ may take any value from set  $\{0,1,2,...,M-1\}$ and there are totally $M$ entities in $\{0,1,2,...,M-1\}$, for one symbol detection, $M$ evaluations of $(r(k) - \hat A (\hat{\mathbf{m}}(k-1,L))m(k))^2$are required. 
Thus, the search complexity still increases with $M$ linearly. 

To reduce the complexity, we can further simplify the decision rule \eqref{eq:dfb1} as
\begin{equation}
\hat{m}(k) =  
\left\{
\begin{array}{rll}
& 0  &, r(k)<0\\
& \lfloor\frac{r(k)}{\hat{A}_k}+\frac{1}{2} \rfloor   &,  \text{elsewhere}\\
&M-1  &, r(k)> (M-1)\hat{A}_k
\end{array}
\right.  .
\label{eq:dfb_rule}
\end{equation}
where  $\hat A_k$ is obtained from \eqref{eq:ch_est_b4sss} and $\lfloor \cdot \rfloor$ is the floor function. 
In contrast to implementing  \eqref{eq:dfb1},  implementing   \eqref{eq:dfb_rule} is much simpler.
First, we need to compare $r(k)$ to 0 and $(M-1) \hat{A}_k$.
If  $r(k) < 0$ or $r(k)>(M-1) \hat{A}_k$, the decision is made. 
If  $0< r(k)<(M-1) \hat{A}_k$, we can substitute $r(k)$ into $\lfloor\frac{r(k)}{ \hat{A}_k}+\frac{1}{2} \rfloor $ and obtain the decision. Totally, the computational complexity is much lower and is independent of the modulation order $M$.

After  implementing   \eqref{eq:dfb_rule}, a reverse Gray mapping is then performed to recovery information bits.

\subsection{The Channel Estimator}

\subsubsection{A selective-store strategy}

If all the elements of the detected subsequence $\hat{\mathbf m}(k-1,L)$ are zeros, the denominator of $\hat A(\hat{\mathbf m}(k-1,L))$ given in \eqref{eq:ch_est_b4sss} is zero, resulting in a channel estimation failure. 
We propose here a \emph{selective store strategy} (SSS): at each time $k$, we only store the $L_m$ most recent received signals that have been detected to carry symbol $M-1$. 
Since all the signals correspond to the same data symbol $M-1$, store of detection results is unnecessary, and thus, the estimate of $A$ reduces to  
\begin{align}
\hat A_k = \frac{\sum_{i=1}^{L_m}r_{i,k}^{M-1}}{ L_m (M-1)},
\label{eq:dfb_h_hat_sss2}
\end{align}
where $r_{i,k}^{M-1}$ is defined as the $i$th most recent received signal at time $k$ that is detected to carry data symbol $M-1$.
In this way, a zero-denominator of the channel estimator will never occur, leading to completely avoiding the channel estimation failure.

\subsubsection{Performance of the DFB receiver with the selective-store strategy}

We assume all the decisions before time $k$ are correct. 
Even in practice, decisions cannot be 100\% correct, with a very low error probability, we can assume so when analysing system performance. 
In the later section that shows simulation results, we do not assume 100\% correct decision feedback.    
Thus, with the assumption, we have $r_{i,k}^{M-1} = A(M-1)+n_{i,k}$, where $n_{i,k}$ is the corresponding AWGN term.
Substituting $r_{i,k}^{M-1} = A(M-1)+n_{i,k}$ into   \eqref{eq:dfb_h_hat_sss2} and simplifying, we have 
\begin{align}
\hat A_k  = A + \frac{\sum_{i=1}^{L_m}n_{i,k}}{L_m (M-1)}.
\end{align}
Obviously, $\hat A_k$ is a Gaussian random variable and its mean and variance are the true value of $A$ and $\frac{N_0}{2 L_m (M-1)^2}$, respectively.
Apparently, we have 
\begin{align} 
  \lim_{L_m\rightarrow \infty} \frac{N_0}{2 L_m (M-1)^2}  = 0.
\end{align}
From \cite[Sec. 14.1]{krishnan2006probability}, we see that for a random variable $X$, if $\mathbb{E}(X-a)^2 = 0$, where $a$ is a constant, the random variable $X$ equals $a$ with probability 1. 
Thus, since  $\text{Var}(\hat h(\hat{\mathbf m}(k-1,L_w))) = \mathbb{E}(\hat h(\hat{\mathbf m}(k-1,L_w))-h)^2$, we have
\begin{align} 
\lim_{L_m\rightarrow \infty} \hat A_k=  A
\label{eq:dfb_hat_h_SSS1} 
\end{align}
with probability 1. 
Since the estimation of $A$ can approach the true value, we can say that the BEP of our DFB receiver with SSS approaches the Genie Bound as $L_m$ goes to infinity.

\subsubsection{The generalised selective-store strategy}

A more general SSS is given as follows: We selectively store the $L_m$ most recent received signals and their corresponding detection results with the criterion $\hat m \geq \alpha $, where $\alpha$ takes value from set $\{1, ..., M-1\}$.
Clearly, $\alpha$ is an important parameter and the SSS introduced in the previous two paragraphs is an extreme case with $\alpha = M-1$. 
If $\alpha \neq M-1$, both detected data and corresponding signals are required to be stored, and the required memory length is $2L_m$; if $\alpha = M-1$, since only signals that are detected to carry $M-1$ are required to be stored, the  memory length is $L_m$ and the best memory efficiency is achieved.
Also, with a fixed value of $L_m$, a higher value of $\alpha$ leads to better performance. 
Another aspect to consider is  to keep the observation window length $L_w$, which is on the order of $L_m M/(M-\alpha)$, far smaller than the channel coherence length $L_c$.   
Thus, to achieve the best performance with a fixed $L_m$ value, we can choose the highest value of $\alpha$ that makes $L_m/(1-\alpha)< L_c/10$.

\section{Performance Results and Discussions}

\subsection{The Genie Bound}

The BEP of symmetric PAM over the AWGN channel has been studied and given in \cite[Eq.'s (9) and (10)]{cho2002general} in terms of the average energy per bit $E_b$ and the AWGN spectral density $N_0$. 
It should be noticed that \cite{cho2002general} assumed an electrical digital communication system and thus $E_b$ is the electrical-domain average energy per bit. 
Based on the mapping given in \cite[Eq. (2)]{cho2002general} which is from $E_b$ to the minimum signal distance $2d$, we can obtain its inverse. 
After substituting this inverse into \cite[Eq.'s (9) and (10)]{cho2002general}, we can derive the general BEP expression in terms of the minimum signal distance $A$ and the AWGN one sided spectral density $N_0$ as  \eqref{eq:bep_pam}. 
\begin{figure*}
 \begin{align}
P& _b^\mathrm{PAM} (\frac{(2d)^2}{N_0},M) =   \sum_{k=1}^{\log_2M} \ \   \sum_{i=0}^{(1-2^{-k})M-1}  
\frac{ (-1)^{\lfloor \frac{i\cdot2^{k-1}}{M}\rfloor} }{M\log_2M} 
    \left( 2^{k} - 2\lfloor \frac{i\cdot 2^{k-1}}{M} + \frac{1}{2} \rfloor   \right) 
  Q \left( (2i+1)\sqrt{\frac{(2d)^2}{2N_0 }}  \right).
\label{eq:bep_pam} 
\end{align} 
\vspace{-30pt}
\end{figure*}
Since \eqref{eq:bep_pam} is expressed in terms $A$ and $N_0$, it applies to both symmetric and asymmetric PAM signals over the AWGN channel. 
Therefore, the BEP of the PCSI receiver conditioned on a given value of $h$ is $P_b^\mathrm{PAM}(\frac{(2hd)^2}{N_0},M)$, and the average BEP over all possible values of $h$ is given by
\begin{align}
P_b^\mathrm{PCSI}(e) = \int_0^\infty P_b^\mathrm{PAM}(\frac{(2hd)^2}{N_0},M) p_h(h)dh .
\label{eq:GenieB} 
\end{align}
This average BEP, which is also referred to as the \emph{Genie Bound}, is used as a benchmark when analysing other receivers.
Specifically, when discussing the BEP of our receiver, we refer to the average BEP over all possible channel states.

{
From \eqref{eq:GenieB}, we see that the average BEP is only related to $2d$, $N_0$, $M$ and $p_h(h)$. 
Thus, when the environment condition, i.e., $p_h(h)$, the system modulation order $M$ and the receiver circuit AWGN one-sided PSD $N_0$ do not change, using different transmission rates (different bandwidths), to achieve the same error probability, we need to keep, $d$ unchanged.  
For conventional electrical digital communication systems, we have
\begin{align}
\frac{1}{M}\sum_{i=0}^{M-1} i^2 (2d)^2 = E_b \log_2M,  
\end{align} 
which is equivalent to
 \begin{align}
2d = \sqrt{\frac{6E_b\log_2M}{(M-1)(2M-1)}},
\label{eq:distance_E_b}
\end{align} 
where $E_b$ denotes the electrical-domain energy per bit.  
From \eqref{eq:distance_E_b}, we see that when keeping the same level of error probability, $E_b$ is irrelevant to the transmission rate.  
This means no matter how fast we transmit data, the energy consumed on each bit does not change, i.e., to transmit the same amount of data, the total energy consumed is irrelevant to the transmission rate.

However, for optical systems, the situation is different. 
The optical-domain energy consumed per bit $E_b^o$ can be calculated via
\begin{align}
E_b^o = \frac{\bar{P}T_s}{\log_2M}. 
\end{align} 
According to \eqref{eq:distance}, we have 
\begin{align}
d^2& = \left( \frac{\sqrt{T_s}R\bar{P}}{M-1} \right)^2  =\frac{ (E_b^o)^2 }{T_s}  \left(\frac{R \log_2M}{M-1}\right)^2  \nonumber \\ 
&= { (E_b^o)^2 } R_\text{data}  \frac{R^2 \log_2M}{(M-1)^2}   ,
\end{align}
where $R_\text{data}=(\log_2M)/T_s$ is defined as the data rate.
We can see that the higher the transmission rate $R_\text{data}$ is, the lower the energy consumed on each bit transmission is. 
Therefore, we can say that, when $2d$, $N_0$, $M$ and $p_h(h)$ are unchanged, i.e., the same level of error probability is retained, to transmit the same amount of data, using a larger data rate (bandwidth) is suggested since the total energy consumption is lower.

This is because the amplitude of the photo detector output current is proportional to the incident light power \cite{govind2010fiber}. 
That means if we increase the optical power by $S$ times, the electrical-domain signal amplitude is increased by $S$ times and the electrical-domain signal power is increased by $S^2$ times. 
When we use a larger bandwidth, say $K$ times of before, to achieve a higher data rate, the noise power, which is the product of the noise power spectral density and the bandwidth, is increased by $K$ times. 
To keep the error probability same, the electrical-domain signal power is also required to be increased by $K$ times, corresponding to increasing the optical signal power by $\sqrt{K}$ times. 
Since the transmission rate is increased by $K$ times, the required time for transmitting the same amount of data is shortened to $\frac{1}{K}$ of before, and thus the total optical energy consumption is $\sqrt{K}\times \frac{1}{K} = \frac{1}{\sqrt{K}}$ times of before. 
Finally, our conclusion is that using a higher bandwidth is suggested since it saves energy.  

It should be emphasized that all the analysis and conclusions in this subsection are based on several basic assumptions: 
\begin{itemize}
\item[1.] pin diode is used for photo detection, i.e., thermal noise can be regarded as the dominant noise souse and shot noise is negligible
\item[2.] the photo detector will never be saturated by the signal pulse that is with the highest power level
\item[3.] the photo detector can response as fast as the user requires, i.e., the photo detector bandwidth is no less than the system symbol rate. 
\item[4.] the FSO channel is always in a linear regime, where the output power is proportional to the input signal power. 
\end{itemize}
Without any of the abovementioned assumptions, the analysis and conclusions may not be true.         
}
\subsection{Simulation Parameters}

 As discussed in \cite{Song2014Arobust} and \cite{Farid2007Outage}, geometric spread and pointing errors $h_p$, atmospheric turbulence $h_a$, and path loss $h_l$  together determine the overall channel gain $h$. 
 The channel state $h$ is formulated as  $h=h_p h_a h_l $.
All these three parameters are time-varying for outdoor environments.
The path loss is related to the atmospheric phenomena, such as rain, fog and snow, and the distance between the transmitter and receiver. 
Therefore, if the distance between the transmitter and the receiver does not chain, $h_l$ changes much slower than $h_a$ and $h_p$. 
In \cite{Song2014Arobust} and \cite{Farid2007Outage}, $h_l$ is considered as a deterministic variable while $h_a$ and $h_p$ are considered as random variable.  
In \cite{Schober2008PCTWC}, log-normal distribution is adopted to model $h_a$ for weak turbulence, Gamma-Gamma distribution for moderate to strong turbulence and the negative exponential distribution for strong turbulence.
Since in \cite{Andrews2001Mathe}, it has been shown that the Gamma-Gamma distribution can nicely fit the channel fading statistics of all turbulence regimes, in this paper, we only consider $h_a$ is a Gamma-Gamma distributed random variable, and the pdf of $h_a$ is  
\begin{align}
p_{h_a}(h)=\frac{2(\alpha\beta)^{(\alpha+\beta)/2}}{\Gamma(\alpha)\Gamma(\beta)}
h^{(\alpha+\beta)/2-1} 
  K_{\alpha-\beta} \left(2\sqrt{\alpha\beta h}\right), 
 h>0 ,  
\label{eq:pdf_gg}
\end{align}
where $K_{a}(\cdot)$ is the modified Bessel function of the second kind, and $1/\beta$ and $1/\alpha$ are the variances of the small and large scale eddies, respectively. 
Pointing error influence on an FSO system is discussed in  \cite{Farid2007Outage} and \cite{Deva2009PointingE}, and we here use the model in \cite{Farid2007Outage} where the pdf of $h_p$ is given as 
\begin{align}
p_{h_p}(h)=\frac{\gamma^2}{A_0^{\gamma^2}} h^{\gamma^2-1} ,\ 0<h<A_0. 
\label{eq: pdf_pointing}
\end{align}
Parameter $A_0$ is the fraction of the collected power when no pointing error occurs, and $\gamma$ is the ratio between the equivalent beam radius at the receiver and the pointing error displacement
standard deviation at the receiver\cite{Farid2007Outage}.
Without loss of generality, we can incorporate $h_l$ into $h_a$ which amounts to setting $h_l=1$. 
Then for a turbulent channel with pointing errors, the channel gain is $h=h_{a} h_p$, and by using \cite[Example 13.1.10]{krishnan2006probability}, its pdf can be derived by  
\begin{align}
p_h(h)=\int_{0}^\infty \frac{1}{|a|}  p_{h_a}(a)p_{h_p}\left(\frac{h}{a}\right) da,\ h>0.
\end{align}
Full model details  can be found in \cite{Song2014Arobust,Schober2008PCTWC, Zhu2002FSOComm, Farid2007Outage,   Andrews2001Mathe,Deva2009PointingE}.

The system parameters of this paper are given as follows: for the weak turbulence channel, $\alpha = 17.13$, $\beta = 16.04$ and the corresponding $\text{SI}=0.1244$; for the strong turbulence channel, $\alpha = 2.23$, $\beta = 1.54$ and $\text{SI}=1.3890$. 
According to\mbox{\cite{andrews2001laserScin}}, the turbulence with SI=0.1244, which is less than 1, is in weak irradiance fluctuations regime; and the turbulence with SI=1.3890, which is larger than 1, is in moderate-to-strong irradiance fluctuations regime.
The pointing error parameters are chosen to be $A_0=0.0198$ and $\gamma = 2.8071$.
Without loss of generality, the photo detector responsivity $R$ is assumed to be 1 and the expected channel gain $\mathbb{E}[h]$ is set to be 1. 
Additionally, we assume the system data rate $R_\text{data}$ is on the order of 10 Gbps, thus the symbol duration $T_s = (\log_2M)/R_\text{data}$ is on the order of $ 10^{-9} - 10^{-10}$s and the channel coherence length $L_c$ can be safely regarded as $10^4$ data symbols. 
We consider the typical thermal noise, which is -174dBm/Hz, passing through a 50$\Omega$ receiver circuit. 
Thus, the value of $N_0/2$ is $ -174\text{dBm/Hz} \div 50\Omega = 10^{-20.4} \text{W/Hz} \div 50\Omega = 7.96\times 10^{-23} \mathrm{A^2/Hz} $, i.e., $N_0 = 1.59\times 10^{-22}\mathrm{A^2/Hz}$.

\subsection{Numerical Results and Discussion}

\begin{figure}[htbp]
\centering
\subfigure[SI = 0.1244.]{  
\includegraphics[scale=0.96]{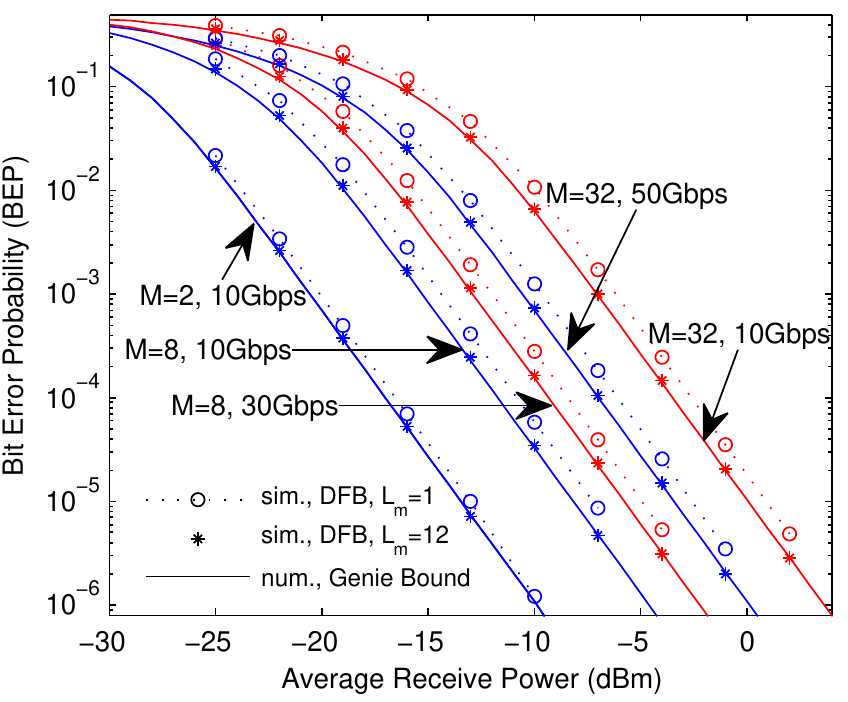}
}

\subfigure[SI=1.3890.]{
\includegraphics[scale=0.96]{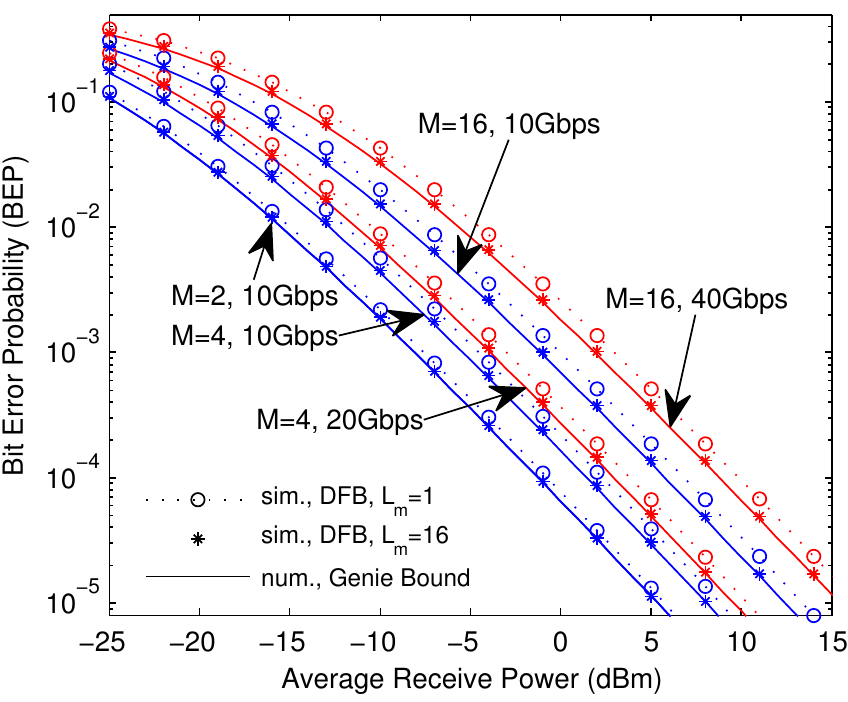}
} 
\caption{Performance of M-PAM signals.}
\label{fg:bep_vs_snr_SSS2}
\vspace{-20pt}
\end{figure}

In Fig. \ref{fg:bep_vs_snr_SSS2}, we plot the BEP versus average receive power curves of our DFB receiver.
Since the implementation complexity of our DFB receiver is independent of $M$, we show simulation results with up to $M=32$. 
We see that as the value of $L_m$ increases, the BEP decreases.
To achieve the Genie Bound, the required value of $L_m$ is 12 for the weak turbulence channel and no larger than 16 for the strong turbulence channel.

From Fig. \ref{fg:bep_vs_snr_SSS2}, we see that if we keep the data rate unchanged and increase $M$ to save bandwidth,  to achieve the same error probability, we need to increase the transmit power. 
This means bandwidth and power are a pair of contradictions. 
We can hardly minimize both of them simultaneously, but we can select appropriate values according to practical requirements.

Another interesting observation is that if we keep $M$ unchanged and use a larger bandwidth to achieve a higher data rate, to achieve the same level error probability, the multiple of the  power growth is the square root of that of the data rate growth.
Specifically, from Fig. \ref{fg:bep_vs_snr_SSS2}, we can see that for $M=4$, if we increase the data rate from 10 Gbps to 20 Gbps, the corresponding power increment is less than 3dB, approximately 1.5 dB.

\begin{figure}
 \centering
 \includegraphics[scale=0.96]{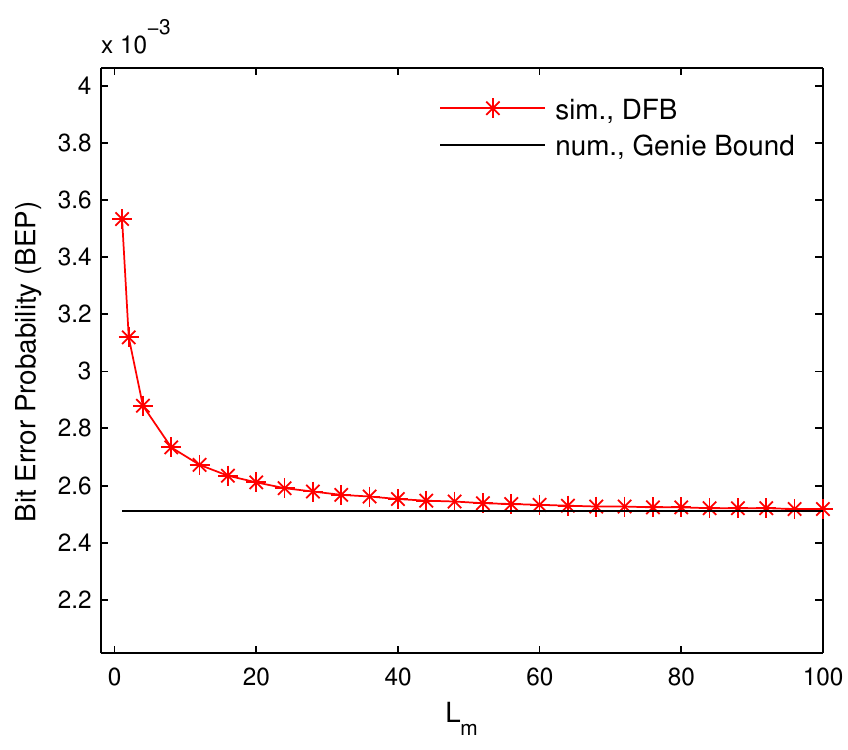}
 \caption{Performance  with different $L_m$'s; $M=16$, -1 dBm, 40 Gbps, SI=1.3890.}
 \label{fg:bep_vs_L}
 \vspace{-10pt}
 \end{figure}

To further study the receiver performance with more values of memory lengths, we plot Fig. \ref{fg:bep_vs_L}. 
For the reason of space limit, we only pick one ($M$, average receive power, data rate, SI) point here, which is ($M=16$, -1 dBm, 40 Gbps, SI=1.3890).
Clearly, as the value of $L_m$ increases, all the BEP curve approach the corresponding Genie Bound.

\begin{figure} 
\centering
\subfigure[$L_m=1$.]{  
\includegraphics[scale=1]{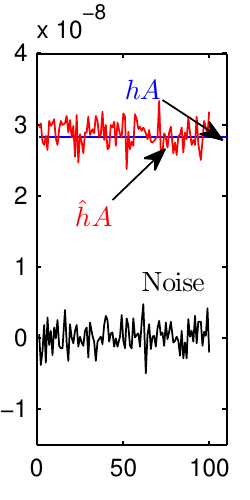}
}
\subfigure[$L_m=4$.]{
\includegraphics[scale=1]{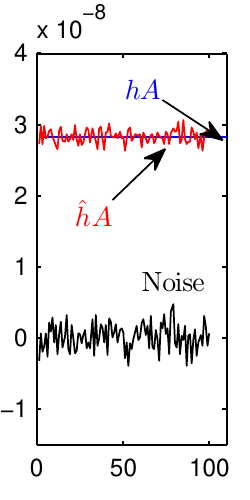}
} 
\subfigure[$L_m=8$.]{
\includegraphics[scale=1]{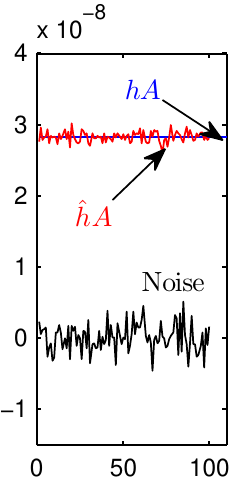}
} 
\caption{Estimates of $I$ with different $L_m$'s; $M=2$, $\frac{A^2}{N_0}=20\text{dB}$.}
\label{fg:estimation}
\vspace{-25pt}
\end{figure}

In Fig. \ref{fg:estimation}, we plot the estimate of $A$ with different memory lengths. 
We assume OOK modulation and choose $\frac{A^2}{N_0}=20\text{dB}$.
The numerical results given in Fig. \ref{fg:estimation} completely agree with the theoretical analysis given in the previous section. 
Thus, we have shown that, by both theoretical analysis and simulation, with a higher value of $L_m$, the estimation of $A$ is more accurate. 
This enables the BEP of our DFB receiver to approach the Genie Bound. 

When starting to operate our DFB receiver, $L_m$ pilot symbols are required to obtain an initial value of $\hat I$. 
Unlike some other communication systems that require frequent insertion of pilot symbols to avoid burst errors, in our simulation, no burst errors are observed even with only $L_m$ pilot symbols for initializing and no further frequent insertions. 
Thus, the pilot-to-data ratio is far smaller than 1, and we do not consider the power consumed by  pilots when calculating the average SNR per bit.

\section{Conclusions}

Since the channel coherence length  $L_c$ is very large, we can use the detected data symbols to estimate the unknown channel state instead of using pilot symbols which causes spectral efficiency reduction. 
In this paper, based on the decision metric of our previously proposed GLRT-MLSD receiver, we propose a DFB symbol-by-symbol receiver, whose implementation complexity  is much lower and is independent of the modulation order.
Hence, it is efficient both spectrally and computationally.  
We also propose a selective-store strategy, which can help avoid potential channel estimation failures and increase the system memory efficiency.
Additionally, we derive a general BEP expression for both symmetric and asymmetric M-PAM signals, by using which as a benchmark, we have shown that as the number of the detected data symbols used to estimate the channel increases, the BEP of our DFB receiver approaches the Genie Bound.

\section*{Acknowledgement}

{  The authors would like to thank the support by the Singapore MoE
AcRF Tier 2 Grant MOE2010-T2-1-101.}


 
\end{document}